\documentclass[aps,prd,onecolumn,nofootinbib]{revtex4-2}
\usepackage{amsmath,amssymb,graphicx}
\usepackage{physics}
\usepackage{hyperref}
\usepackage{esvect}

\begin{document}

\title{Still The New Classical Relativistic Equation of Charge Motion in an Electromagnetic Field}

\author{
    Anatoliy V. Sermyagin \\
   JSC "Institute in Physical-Technical Problems",\\ Dubna, Russia, \\
    sermyagin@jinr.ru
}

\date{\today}

\begin{abstract}
The non-relativistic Goedecke equation~(1975), which describes the motion of a point charge taking into account the radiation reaction, has no ``runaway'' solutions. A ``physical'' method of covariant generalization of this equation is proposed, a special case of which is based on the Lorentz transformations in a coordinate--free covariant representation. Two equivalent forms of a new classical relativistic equation of motion of a point charge are obtained. It is shown that the Abraham--Lorentz--Dirac (ALD) and the Mo--Papas (MP) equations are approximate consequences of the presented theory.
\end{abstract}

\maketitle

\vspace{1em}

\section{Preface}
This paper is based upon the materials of my old preprint from 1978, which has undergone, on my part, a thorough incite and revision. Therefore, despite the fact that a lot of time has passed since the first publication, I considered it possible to call this work ``Still the new classical relativistic...''.
\section{Covariantization the retarded classical equation}\label{sec:constraints}
Let us consider the non-relativistic equation of motion of a point charge $ e $ with mass $ m $ in an external electromagnetic field $ \vv{\bf E}  $, $ \vv{\bf H} $ \cite{Goedecke1975}: 
\begin{equation}
{{\ddot {\vv {\bf r}}}}\left( {t - {\tau _0}} \right) = 
\frac{e}{m}\left\{ \vv{\bf E} + \left[ \dot {\vv {\bf r}}\left( t \right) {\vv{\bf H}} \right] \right\}, 
\label{eq:1}
\end{equation}
where $ {\tau _0} = \frac{2}{3}\frac{{{e^2}}}{m} $, and the speed of light $ c = 1 $ . The dot above the symbol indicates the time derivative, $ \dv*{{\vv {\bf r}}\left( t \right)}{t} \equiv  {{\dot {\vv {\bf r}}}}\left( {t } \right) $.
The problem of the electromagnetic mass of a point charge is not considered here; it is assumed that the mass has been renormalized in any suitable way.

Equation (\ref{eq:1}) can easily be reduced to the non-relativistic Abraham-Lorentz (AL) equation, \cite{Jack}, which describes the motion of a charge under the action of an external electromagnetic field and the reaction of its proper electromagnetic field, if we expand the left-hand side of (\ref{eq:1}) in a series in the small parameter ${\tau _0}$ and restrict ourselves to the linear in ${\tau _0}$ term of this expansion.

A special feature of equation (\ref{eq:1}), unlike the AL equation, is the absence of runaway solutions, but there is a pre-acceleration effect which is not the subject of our consideration here.

The point is to find a relativistic generalization of equation (\ref{eq:1}) for which it would be convenient utilize a 4-dimensional notation
(see \ref{sec:A}). However, direct replacement of 3-dimensional spatial vectors $ {{\dot {\vv {\bf r}}}}\left( {t } \right) $, $ {{\ddot {\vv {\bf r}}}}\left( {t- {\tau _0} } \right) $, $ \vv{\bf E} $, $ \vv{\bf H} $ with their 4-dimensional counterparts $ {v^i}(\tau ) $, $ {{\dot v}^i}(\tau  - {\tau _0}) $, and the tensor of the external electromagnetic field  $ {F^{ij}}(\tau ) $, respectively, does not immediately lead to the correct result. Indeed, the left and right parts of equation (\ref{eq:1}) in covariant notation have the form:
\begin{equation}
{{\dot v}^i}(\tau  - {\tau _0}) 
\label{eq:2}
\end{equation}

\begin{equation}
\frac{e}{m}{F^{ij}}(\tau ){v_j}(\tau ) 
\label{eq:3}
\end{equation}
The scalar product of the 4-Lorentz force Eq.~(\ref{eq:3}) by the 4-velocity $ {v^i}(\tau ) $ gives an identical zero due to the skew symmetry of the electromagnetic field tensor with respect to rearranging indexes, $ {F^{ij}}(\tau ) =  - {F^{ji}}(\tau ) $:
\begin{equation}
\frac{e}{m}{F^{ij}}(\tau ){v_j}(\tau ){v_i}(\tau ) \equiv 0. 
\label{eq:4}
\end{equation}
But the result of multiplying ~(\ref{eq:2})  by the 4-velocity vector $ {v_i}(\tau ) $  is not identically zero, in the general case,
\begin{equation}
{{\dot v}^i}(\tau  - {\tau _0}){v_i}(\tau ) \not\equiv 0, 
\label{eq:5}
\end{equation}
since the 4-speed of the charge is $ {{v}^i}(\tau)\equiv \dv*{x^i}{\tau} $ and 4-acceleration of the charge $ {\dot{v}^i}(\tau  - {\tau _0}) \equiv \dv*{{v^i}(\tau  - {\tau _0})}{\tau} $, taken at different moments of their proper time  $ \tau  - {\tau _0} $  and $ \tau $, refer to different inertial instantaneous-accompanying reference frames, and, therefore, do not always have to be orthogonal to each other. Here, $ \mathrm{d}\tau  = \mathrm{d}t \sqrt {1 - {{\left( \dv*{{\vv {\bf r}}(t)}{t}  \right)}^2}} $, and $ {\vv {\bf r}}(t) $ be the spatial radius vector of the charge position.
One of the ways to obtain a relativistic analogue of an equation of type (\ref{eq:1}) is to perform the orthogonalization procedure on the 4-vectors $ {v^i}(\tau ) $, $ \dot{v}^i(\tau - \tau_0 ) $, like obtaining the covariant ALD equation, \cite{dirac1938}, from the AL equation, \cite{LL72}. To do this, to the 4-vector $ {{\dot v}^i}(\tau  - {\tau _0}) $  we add the 4-vector
\begin{equation}
 - {\dot v^k}\left( {\tau  - {\tau _0}} \right){v_k}\left( \tau  \right){v^i}\left( \tau  \right), 
\label{eq:6}
\end{equation}
after that, the left side of the equation under constraction takes the form
\begin{equation}
 {\dot v^i}\left( {\tau  - {\tau _0}} \right) - {\dot v^k}\left( {\tau  - {\tau _0}} \right){v_k}\left( \tau  \right){v^i}\left( \tau  \right). 
\label{eq:7}
\end{equation}
In other words, the 4-vector $ {\dot v^i}\left( {\tau  - {\tau _0}} \right) $ be multiplied by the orthogonal projector
\begin{equation}
{\rm  {\delta}}_j^i - {v^i}\left( \tau \right){v_j}\left( \tau  \right).  
\label{eq:8}
\end{equation}
As a result, equation \cite{Sorg} follows from (\ref{eq:7}):
\begin{equation}
{{\dot v}^i}(\tau  - {\tau _0}) - {{\dot v}^k}(\tau  - {\tau _0}){v_k}(\tau ){v^i}(\tau ) = \frac{e}{m}{F^{ij}}(\tau ){v_j}(\tau )
\label{eq:9}
\end{equation}
It is easy to check the orthogonality of both parts of equation (\ref{eq:9}) to the 4-velocity $ {v^i}(\tau ) $ directly.
However, this method of orthogonalization has the disadvantage that it is difficult to give it a physical meaning, since equation (\ref{eq:9}) is not a strictly covariant generalization of equation (\ref{eq:1}) - orthogonal projection does not preserve the norm of the 4-vector $ {{\dot v}^i}(\tau  - {\tau _0}) $. 
In \cite{AVS1978}, another method was proposed that allows us to talk about a consistent relativistic approach to obtaining covariant generalizations of non-relativistic delayed-motion dynamics equations. The physical reason that $ {{v}^i}(\tau) $   and $ {{\dot v}^i}(\tau  - {\tau _0}) $  are not orthogonal is that they refer, generally speaking, to two different instantaneously accompanying frames of reference moving in an inertial laboratory the frame of reference an observer with different 4-velocities, $ {v^i}(\tau) \not \equiv {v^i}(\tau  - {\tau _0}) $. 
Therefore, the ``physical'' procedure for orthogonalizing the quantities $ {{ v}^i}(\tau) $ and $ {{\dot{v}}^i}(\tau  - {\tau _0}) $ consists in reducing these quantities to the same inertial instantaneous frame of reference, for that it is necessary to perform the Lorentz transformation $ \Lambda \left( {u \Rightarrow v} \right) $ 4-acceleration vectors $ {{\dot v}^i}(\tau  - {\tau _0}) $  from the reference frame, 4-velocity of which in the laboratory reference frame is $ {{v}^i}(\tau  - {\tau _0}) \equiv u $, to the reference frame whose 4-velocity is $ {v^i}(\tau) \equiv v $ in the laboratory reference frame. As result of this physical orthogonalization, the correct covariant relativistic generalization of equation (\ref{eq:1}) takes the form:
\begin{equation}
\Lambda _k^i{{\dot v}^k}\left( {\tau  - {\tau _0}} \right) = \frac{e}{m}{F^{ik}}{v_k}\left( \tau  \right),
\label{eq:10}
\end{equation}
where the Lorentz transformation $ \Lambda _k^i $  is defined as 
$ \Lambda _k^i{{v}^k}(\tau  - {\tau _0}) = {v^i}(\tau) $. 
We should immediately note that due to the fact that
\begin{equation}
{\Lambda _k^i}{{\Lambda ^{ - 1}}_j^k} = \rm{I}_j^i,
\label{eq:11}
\end{equation}
where $ {{\Lambda ^{ - 1}}_j^k} \Rightarrow \Lambda ^{ - 1} \cdot v = u $ is the  inverse of the transformation $ \Lambda $, and $ \rm{I}_j^i $  is the identical transformation,
\begin{equation}
\rm{I}_j^i \equiv \left\{ {\begin{array}{*{10}{c}}
{1,i = j}\\
{0,i \ne j}
\end{array}} \right.,
\label{eq:12}
\end{equation}
the equation
\begin{equation}
{{\dot v}^i}\left( {\tau  - {\tau _0}} \right) = \frac{e}{m}{\Lambda ^{ - 1}}_j^i{F^{jk}}\left( \tau  \right){v_k}\left( \tau  \right) 
\label{eq:13}                                   
\end{equation}
be equivalent to (\ref{eq:10}).

Using index-free notation (\ref{sec:A}) and explicit expressions for 
the Lorentz transformations $ \Lambda $  and $ \Lambda^{-1} $, the derivation of which is given in (\ref{sec:AB}), from (\ref{eq:33}) and (\ref{eq:34}) we obtain a relativistic generalization of equation (\ref{eq:1}) in two equivalent forms:
\begin{equation}
m\dot u - m\dot u \cdot v\frac{{u + v}}{{1 + u \cdot v}} = eF \cdot v,
\label{eq:14}
\end{equation}
\begin{equation}
m\dot u = eF \cdot v - eF \cdot v \cdot u\frac{{u + v}}{{1 + u \cdot v}},
\label{eq:15}
\end{equation}
where $ u \equiv v\left( {\tau  - {\tau _0}} \right) $, $ v \equiv v\left( \tau  \right) $, ${\tau _0} = \frac{2}{3}\frac{{{e^2}}}{m} $, $ c = 1 $.
It is important to note that the Lorentz transformations used to derive (\ref{eq:14}) and (\ref{eq:15}) do not contain spatial rotations. Therefore, strictly speaking, (\ref{eq:14}) and (\ref{eq:15}) describe only one-dimensional motions, while equations (\ref{eq:10}) and (\ref{eq:13}) are applicable to the general situation. An adequate covariant representation of such general Lorentz transformations will be published separately [].

The fact that equations (\ref{eq:14}) and (\ref{eq:15}) do not have “runaway” solutions for $ F = 0 $  is evident from the general form of these equations (\ref{eq:10}) and (\ref{eq:13}), when the representations of the operators $ \Lambda $  and $ \Lambda^{-1} $  be not specified. For (\ref{eq:13}) and (\ref{eq:15}), this is obvious; under the condition $ F = 0 $ , we act on the left and right sides of equations (\ref{eq:10}) and (\ref{eq:14}) with the operator  $ \Lambda^{-1} $ , we will have then $ \dot u = 0 $, i.e., the equation of motion of a free particle.

Note the connection of (\ref{eq:14}) and (\ref{eq:15}) with the ALD and MP equations, which be approximate consequences of the presented theory. To do this, we decompose the quantities depending on the lagging argument $ \tau  - {\tau _0} $  by degrees of delay $ {\tau _0} $  and discard the terms proportional to $ {\tau _0} $  to a power higher than the first. We will then obtain from (\ref{eq:14}) the equation ALD,
\begin{equation}
\dot v - {\tau _0}\left( {\ddot v + \dot v \cdot \dot vv} \right) = \frac{e}{m}F \cdot v.
\label{eq:16}
\end{equation}
In this case, it follows from (\ref{eq:15})
\begin{equation}
\dot v - {\tau _0}\ddot v = \frac{e}{m}F \cdot v + {\tau _0}\frac{e}{m}F \cdot v \cdot \dot vv.
\label{eq:17}
\end{equation}
Using the ratio $ \ddot v = \frac{e}{m}\dv*{(F \cdot v)}{\tau} $, assuming that $ \dot F \cdot v = 0 $, we obtain from (\ref{eq:17}) the equation MP \cite{MoP}:
\begin{equation}
\dot v = \frac{e}{m}F \cdot v + {\tau _0}\frac{e}{m}\left( {F \cdot \dot v + F \cdot v \cdot \dot vv} \right).
\label{eq:18}
\end{equation}
For the first time, equations (\ref{eq:7}), (\ref{eq:10}), (\ref{eq:14}) and (\ref{eq:15}) were presented in \cite{AVS1978}. The investigation of solutions to these equations will be published separately.

\section{Discussion and Conclusion} \label{sec:discussion}

We have achieved the stated goal of obtaining a covariant generalization of equation (\ref{eq:1}) through proper physical covariantization and got the new relativistic equation in two equivalent forms, (\ref{eq:10}), (\ref{eq:13}), (\ref{eq:14}) and (\ref{eq:15}).  Since the Lorentz transformations used do not contain spatial rotations, the resulting equations (\ref{eq:14}) and (\ref{eq:15}) be of a special case describing one-dimensional motion in the external electromagnetic field. An analysis of general equations (\ref{eq:10}) and (\ref{eq:13}) will be published later.
\\
\\

\section{\bf Appendix A: Coordinate\&Index-Free Notation and Normalizing}\label{sec:A}

As usual, the summation rule is used (without explicitly specifying the sign of the sum $ \sum $) for repeated indexes that take values 0,1,2,3:
\begin{equation}
{x^k}{y_k} \equiv \sum\limits_{i,j = 0}^3 {{x^i}{y^j}{\eta _{ij}}}, 
\label{eq:19}
\end{equation}
where $ \eta _{ij} $ denotes the metric tensor of the Minkowski space,
\begin{equation}
{\eta _{ij}} = \left[ {\begin{array}{*{20}{c}}
1&0&0&0\\
0&{ - 1}&0&0\\
0&0&{ - 1}&0\\
0&0&0&{ - 1}
\end{array}} \right].
\label{eq:20}
\end{equation}
Where not stated otherwise, the 4-velocity vectors are denoted by small Latin letters with no indexes and no indicated dependence on the proper time $ \tau $,  $ u \equiv v\left( {\tau  - {\tau _0}} \right) $, $ v \equiv v\left( \tau  \right) $. Inner (scalar) product of pair 4-vectors is indicated with a dot between these vectors; in the case of 4-vectors: $ x^k y_k \Rightarrow x \cdot y $, in the case of a second rank tensor and 4-vector: $ F^{ ik } y_k \equiv F^ i_k v^k \equiv  F^{ ik } \eta _{kj} v^j \Rightarrow F \cdot v $. 4-speed normalization:  
\begin{equation}
 v \cdot v = 1.
\label{eq:21}
\end{equation}
The orthogonality of 4-velocity and 4-acceleration is a consequence of normalization (\ref{eq:21}). Indeed, differentiating (\ref{eq:21}) by proper time $ \tau $, we obtain:
\begin{equation}
 \frac{\rm d}{{\rm d\tau }}v \cdot v = 2\dot v \cdot v \equiv 0. 
 \label{eq:22}
\end{equation}

\section{\bf Appendix B: Covariantization the Lorentz Transformation}\label{sec:AB}
The possibility of factoring Lorentz transformations between two inertial reference frames, in the formalism of tetrad, that is, 4-dimensional basis vectors, was established by Bazansky \cite{Baz1965}. An explicit form of Lorentz transformations, without spatial rotations, as functions of 4-velocities in index-free notation, was obtained by J.~Krause \cite{JK1977,JK1978}. Such Lorentz transformations, derived independently in a different way, are used in the author's work \cite{AVS1978} for a covariant generalization equation (\ref{eq:1}).

If $ \Lambda $ is a homogeneous or proper Lorentz transformation without spatial rotations, $ \Lambda :u \Rightarrow v $, so that
\begin{equation}
\cosh \varphi = u \cdot v,
\label{eq:23}
\end{equation}
then the matrix of $ {\bf{M}} $ rotations in the 1+3 Minkowski space corresponds to this transformation,
\begin{equation}
{\bf{M}} = \left[ {\begin{array}{*{20}{c}}
{\cosh \varphi }&{\sinh \varphi }&0&0\\
{\sinh \varphi }&{\cosh \varphi }&0&0\\
0&0&{1}&0\\
0&0&0&{1}
\end{array}} \right],
\label{eq:24}
\end{equation}
with respect to the orthonormal local basis $ \varepsilon $,
\begin{equation}
\varepsilon  = \left[ {\begin{array}{*{20}{c}}
{{\varepsilon _0}}\\
{{\varepsilon _1}}\\
{{\varepsilon _2}}\\
{{\varepsilon _3}}
\end{array}} \right],  {\varepsilon ^T} = \left[ {\begin{array}{*{20}{c}}
{{\varepsilon ^0}}&{{\varepsilon ^1}}&{{\varepsilon ^2}}&{{\varepsilon ^3}}
\end{array}} \right],                                   
\label{eq:25}
\end{equation}
$ {{\varepsilon _0} = {\varepsilon ^0} \equiv u} $, $ {{ \varepsilon _1} = {\varepsilon ^1} \equiv \frac{v}{{\sinh \varphi }} - u\cosh \varphi } $, 
``$ ...^T $'' means transposition.   
We don't need exact expressions for $ \varepsilon _2 $ and $ \varepsilon _3 $ here. 
An arbitrary 4-vector $ a $ can be written as a basis vector expansion (\ref{eq:25}),
\begin{equation}
a \equiv {\varepsilon ^T} {\bf{G}}a \cdot  \varepsilon, 
\label{eq:26}
\end{equation}
The matrix $ \bf{G} $ is defined by the relation
\begin{equation}
{\bf{G}} \equiv {\eta _{ij}} = \varepsilon {\varepsilon ^T}.
\label{eq:27}
\end{equation}
Now, using decomposition (\ref{eq:26}), we write the Lorentz transformation of the 4-vector $ a $ in general form as:
\begin{eqnarray} 
\begin{split}
 b &= \Lambda a = {\varepsilon ^T}   {\bf{MG}}a \cdot  \varepsilon  \\
 &= a - a + {\varepsilon ^T}   {\bf{MG}}a \cdot \varepsilon   \\
 &= a + {\varepsilon ^T}   ({\bf{M}} - {\bf{I}}){\bf{G}}a \cdot \varepsilon,   
\end{split}
\label{eq:28} 
\end{eqnarray}
where $ {\bf{I}} $  is the identity matrix,
\begin{equation}
{\bf{I}} = \left[ {\begin{array}{*{20}{c}}
1&0&0&0\\
0&1&0&0\\
0&0&1&0\\
0&0&0&1
\end{array}} \right]
\label{eq:29}
\end{equation}
We also use light font to denote identity matrix (\ref{eq:29}), $ {\bf I} \Rightarrow  {\rm I}  $.
From (\ref{eq:28}), we obtain the Lorentz transformation 
in a coordinate-free covariant form:
\begin{eqnarray}
\begin{split}
{\Lambda}&={ {\varepsilon ^T}{\bf{MG}}\varepsilon} 
= {\bf{I} + {\varepsilon ^T} ({\bf{M}} - {\bf{I}}){\bf{G}}\varepsilon }
= \left[ {\begin{array}{*{20}{c}}
1&0&0&0\\
0&1&0&0\\
0&0&1&0\\
0&0&0&1
\end{array}} \right] + \left[ {\begin{array}{*{20}{c}}
u&{u\cosh \varphi  - \frac{v}{{\sinh \varphi }}}&{{\varepsilon ^2}}&{{\varepsilon ^3}}
\end{array}} \right] \times \\ 
&\times\left\{ {\left[ {\begin{array}{*{20}{c}}
{\cosh \varphi }&{\sinh \varphi }&0&0\\
{\sinh \varphi }&{\cosh \varphi }&0&0\\
0&0&{1}&0\\
0&0&0&{1}
\end{array}} \right]} 
{-} \left[ \begin{array}{*{20}{c}}
1&0&0&0\\
0&1&0&0\\
0&0&1&0\\
0&0&0&1
\end{array} \right]\right\}
 \left[ {\begin{array}{*{20}{c}}
1&0&0&0\\
0&{-1}&0&0\\
0&0&{-1}&0\\
0&0&0&{-1}
\end{array}} \right] 
\left[ {\begin{array}{*{20}{c}}
u\\
{u\cosh \varphi  - \frac{v}{{\sinh \varphi }}}\\
{{\varepsilon _2}}\\
{{\varepsilon _3}}
\end{array}} \right] \\
&={\rm{I}} + \left[ {\begin{array}{*{20}{c}}
u&{u\cosh \varphi  - \frac{v}{{\sinh \varphi }}}&{{\varepsilon ^2}}&{{\varepsilon ^3}}
\end{array}} \right] \times {\left[ {\begin{array}{*{20}{c}}
{\cosh \varphi - 1 }&{\sinh \varphi }&0&0\\
{\sinh \varphi }&{\cosh \varphi - 1}&0&0\\
0&0&{0}&0\\
0&0&0&{0}
\end{array}} \right]} 
\left[ {\begin{array}{*{20}{c}}
u\\
{u\cosh \varphi  - \frac{v}{{\sinh \varphi }}}\\
{{\varepsilon _2}}\\
{{\varepsilon _3}}
\end{array}} \right] \\
&={\rm{I}} + \left[ {\begin{array}{*{20}{c}}
u&{u\cosh \varphi  - \frac{v}{{\sinh \varphi }}}
\end{array}} \right]\left[ {\begin{array}{*{20}{c}}
{\cosh \varphi  - 1}&{\sinh \varphi }\\
{\sinh \varphi }&{\cosh \varphi  - 1}
\end{array}} \right]
\left[ {\begin{array}{*{20}{c}} u\\
{ - u\cosh \varphi  + \frac{v}{{\sinh \varphi }}}
\end{array}} \right].
\end{split}
\label{eq:30}
\end{eqnarray}

Performing matrix product in (\ref{eq:30}), remembering that $ \cosh \varphi = u \cdot v $, we get:
\begin{equation}
\Lambda  = {\rm{I}} - u\frac{{u + v}}{{1 + u \cdot v}} + v\frac{{2u \cdot vu + u - v}}{{1 + u \cdot v}}.
\label{eq:31}
\end{equation}
The inverse transformation has the form:
\begin{equation}
 {\Lambda ^{ - 1}} = {\varepsilon ^T}{{\bf{M}}^{ - 1}}{\bf{G}}\varepsilon  = {\bf{I}} + {\varepsilon ^T} \cdot ({{\bf{M}}^{ - 1}} - {\bf{I}}){\bf{G}}\varepsilon, 
\label{eq:32}
\end{equation}
\begin{equation}
 {{\bf{M}}^{ - 1}} = \left[ {\begin{array}{*{20}{c}}
{\cosh \varphi }&{ - \sinh \varphi }&0&0\\
{ - \sinh \varphi }&{\cosh \varphi }&0&0\\
0&0&1&0\\
0&0&0&1
\end{array}} \right],
\label{eq:33}
\end{equation}
\begin{equation}
\begin{split}
{\Lambda ^{ - 1}} &= \left[ {\begin{array}{*{20}{c}}
1&0&0&0\\
0&1&0&0\\
0&0&1&0\\
0&0&0&1
\end{array}} \right] + \left[ {\begin{array}{*{20}{c}}
u&{u\cosh \varphi  - \frac{v}{{\sinh \varphi }}}&{{\varepsilon ^2}}&{{\varepsilon ^3}}
\end{array}} \right]\times\\
&\times\left\{ {\left[ {\begin{array}{*{20}{c}}
{\cosh \varphi }&{ - \sinh \varphi }&0&0\\
{ - \sinh \varphi }&{\cosh \varphi }&0&0\\
0&0&1&0\\
0&0&0&1
\end{array}} \right]} \right. - \left. {\left[ {\begin{array}{*{20}{c}}
1&0&0&0\\
0&1&0&0\\
0&0&1&0\\
0&0&0&1
\end{array}} \right]} \right\}\times \\ 
&\times\left[ {\begin{array}{*{20}{c}}
1&0&0&0\\
0&{ - 1}&0&0\\
0&0&{ - 1}&0\\
0&0&0&{ - 1}
\end{array}} \right]\left[ {\begin{array}{*{20}{c}}
u\\
{u\cosh \varphi  - \frac{v}{{\sinh \varphi }}}\\
{{\varepsilon _2}}\\
{{\varepsilon _3}}
\end{array}} \right]\\
&={\rm{I}} + \left[ {\begin{array}{*{20}{c}}
u&{u\cosh \varphi  - \frac{v}{{\sinh \varphi }}}
\end{array}} \right]\left[ {\begin{array}{*{20}{c}}
{\cosh \varphi  - 1}&{-\sinh \varphi }\\
{-\sinh \varphi }&{\cosh \varphi  - 1}
\end{array}} \right]
\left[ {\begin{array}{*{20}{c}} u\\
{ - u\cosh \varphi  + \frac{v}{{\sinh \varphi }}}
\end{array}} \right]\\
&={\rm{I}} - v\frac{{u + v}}{{1 + u \cdot v}} + u\frac{{2u \cdot vv + v - u}}{{1 + u \cdot v}}.
\label{eq:34}
\end{split}
\end{equation}
The forward ${\Lambda}$ and inverse ${\Lambda ^{ - 1}}$  Lorentz transformations differ by permutation of 4-velocities, which is obvious from a comparison of their form:
\begin{equation}
\Lambda  = {\rm{I}} - u\frac{{u + v}}{{1 + u \cdot v}} + v\frac{{2u \cdot vu + u - v}}{{1 + u \cdot v}},
\label{eq:35}
\end{equation}
\begin{equation}
{\Lambda ^{ - 1}} = {\rm{I}} - v\frac{{u + v}}{{1 + u \cdot v}} + u\frac{{2u \cdot vv + v - u}}{{1 + u \cdot v}}.
\label{eq:36}
\end{equation}
We also note that
\begin{equation}
\Lambda \cdot u = v, v \cdot \Lambda  = u, {\Lambda ^{ - 1}} \cdot v = u, u \cdot {\Lambda ^{ - 1}} = v. 
\label{eq:37}
\end{equation}                     
When using a coordinate-free/index-free representation, it is necessary to keep in mind the matrix nature of the multiplied quantities and their noncommutativity. In this regard, a modified 4-vector representation formalism may be convenient, in which the Lorentz transformations look like this:
\begin{equation}
\Lambda  = {\rm{I}} - \left. {|u} \right)\frac{{\left( {u|} \right. + \left( {v|} \right.}}{{1 + \left( {u|v} \right)}} + \left. {|v} \right)\frac{{2\left( {u|v} \right)\left( {u|} \right. + \left( {u|} \right. - \left( {v|} \right.}}{{1 + \left( {u|v} \right)}},
\label{eq:38}
\end{equation}
\begin{equation}
 {\Lambda ^{ - 1}} = {\rm{I}} - \left. {|v} \right)\frac{{\left( {u|} \right. + \left( {v|} \right.}}{{1 + \left( {u|v} \right)}} + \left. {|u} \right)\frac{{2\left( {u|v} \right)\left( {v|} \right. + \left( {v|} \right. - \left( {u|} \right.}}{{1 + \left( {u|v} \right)}}.
\label{eq:39}
\end{equation}
An arbitrary 4-vector $ p $ may be written in bracket notation as:
\begin{equation}
{p_k} \equiv \left( {p|} \right., {p^i} \equiv \left. {|p} \right).
\label{eq:40}  
\end{equation}
i.e., the transition between bra- and -cket vectors corresponds to the procedure of raising and lowering indices in index notation using the Minkovsky metric tensor, (\ref{eq:20}).
The scalar product of 4-vectors $ a $ and $ b $: 
\begin{equation}
{a^i}{b_i} \Rightarrow a \cdot b \equiv b \cdot a \equiv \left( {a|b} \right) \equiv \left( {b|a} \right),
\label{eq:41}
\end{equation}
tensor product of 4-vectors $ a $ and $ b $:
\begin{equation}
{a^i}{b_k} \Rightarrow ab \equiv \left. {|a} \right)\left( {b|} \right..  
\label{eq:42}
\end{equation}
In this entry, the Lorentz transformation $ \Lambda :u \Rightarrow v $ looks like this:
\begin{equation}
\begin{split}
{\Lambda  \cdot u} &\Rightarrow \left\{ {{\rm{I}} - \left. {|u} \right)\frac{{\left( {u|} \right. + \left( {v|} \right.}}{{1 + \left( {u|v} \right)}} + \left. {|v} \right)\frac{{2\left( {u|v} \right)\left( {u|} \right. + \left( {u|} \right. - \left( {v|} \right.}}{{1 + \left( {u|v} \right)}}} \right\}\left. {|u} \right)\\
&= \left. {|u} \right) - \left. {|u} \right)\frac{{\left( {u|u} \right) + \left( {v|u} \right)}}{{1 + \left( {u|v} \right)}} + \left. {|v} \right)\frac{{2\left( {u|v} \right)\left( {u|u} \right) + \left( {u|u} \right) - \left( {v|u} \right)}}{{1 + \left( {u|v} \right)}}\\
&= \left. {|u} \right) - \left. {|u} \right) + \left. {|v} \right)\\
&= \left. {|v} \right).
\end{split}  
\label{eq:43}
\end{equation}
Similarly, it is easy to prove that $ {\Lambda ^{ - 1}} \cdot v = u $. The properties of covariant Lorentz transformations can be found in \cite{JK1977,JK1978}.  
Substituting (\ref{eq:38}) into (\ref{eq:10}), taking into account (\ref{eq:37}), we get:
\begin{equation}
\begin{split}
m{\Lambda}_j^i{\dot u^{j}} &\Rightarrow m\Lambda  \cdot \dot u \Rightarrow m\left\{ {{\rm{I}} - \left. {|u} \right)\frac{{\left( {u|} \right. + \left( {v|} \right.}}{{1 + \left( {u|v} \right)}} + \left. {|v} \right)\frac{{2\left( {u|v} \right)\left( {u|} \right. + \left( {u|} \right. - \left( {v|} \right.}}{{1 + \left( {u|v} \right)}}} \right\}\left. {|\dot u} \right)\\
 &= m\left\{ {\left. {|\dot u} \right) - \left. {|u} \right)\frac{{\left( {u|\dot u} \right) + \left( {v|\dot u} \right)}}{{1 + \left( {u|v} \right)}} + \left. {|v} \right)\frac{{2\left( {u|v} \right)\left( {u|\dot u} \right) + \left( {u|\dot u} \right) - \left( {v|\dot u} \right)}}{{1 + \left( {u|v} \right)}}} \right\}\\
 &= m\left. {|\dot u} \right) - m\left( {v|\dot u} \right)\frac{{\left. {|u} \right) + \left. {|v} \right)}}{{1 + \left( {u|v} \right)}}\\
 &\Rightarrow m\dot u - m\dot u \cdot v\frac{{u + v}}{{1 + u \cdot v}} = eF \cdot v.
\end{split}  
\label{eq:44}
\end{equation}
For the Lorentz transformation the Lorentz force, we note that
\begin{equation}
\begin{split}
{F^{ik}} &=  - {F^{ki}};\\
{F^{ik}}{v_i}{v_k} = {F^{ik}}{u_i}{u_k} & \Rightarrow F \cdot u \cdot u = F \cdot v \cdot v \equiv 0;\\
{F^{ik}}{v_k} &\Rightarrow F \cdot v \Rightarrow \left. {|F \cdot v}\right).
\end{split}
\label{eq:45}
\end{equation}
Substituting (\ref{eq:39}) into (\ref{eq:13}), taking into account (\ref{eq:45}), we obtain:
\begin{equation}
\begin{split}
{\Lambda ^{ - 1}}_j^i{F^{jk}}v &\Rightarrow {\Lambda ^{ - 1}}\left. {|F \cdot v} \right)
 \equiv \left\{ {{\rm{I}} - \left. {|v} \right)\frac{{\left( {u|} \right. + \left( {v|} \right.}}{{1 + \left( {u|v} \right)}} + \left. {|u} \right)\frac{{2\left( {u|v} \right)\left( {v|} \right. + \left( {v|} \right. - \left( {u|} \right.}}{{1 + \left( {u|v} \right)}}} \right\}\left. {|F \cdot v} \right)\\
 &= \left. {|F \cdot v} \right) - \left. {|v} \right)\frac{{\left( {u|F \cdot v} \right) + \left( {v|F \cdot v} \right)}}{{1 + \left( {u|v} \right)}} + \left. {|u} \right)\frac{{2\left( {u|v} \right)\left( {v|F \cdot v} \right) + \left( {v|F \cdot v} \right) - \left( {u|F \cdot v} \right)}}{{1 + \left( {u|v} \right)}}\\
 &= \left. {|F \cdot v} \right) - \left. {|v} \right)\frac{{\left( {u|F \cdot v} \right)}}{{1 + \left( {u|v} \right)}} - \left. {|u} \right)\frac{{\left( {u|F \cdot v} \right)}}{{1 + \left( {u|v} \right)}}\\
 &= \left. {|F \cdot v} \right) - \frac{{\left. {|u} \right) + \left. {|v} \right)}}{{1 + \left( {u|v} \right)}}\left( {u|F \cdot v} \right).
\end{split}
\label{eq:46}
\end{equation}
As a result, the second equivalent form of the covariant generalization of equation (\ref{eq:1}) in index-free notation takes the form:
\begin{equation}
m\dot u = eF \cdot v - eF \cdot v \cdot u\frac{{u + v}}{{1 + u \cdot v}}.
\label{eq:47}
\end{equation}


\subsection*{Acknowledgment}
I would like to thank Dr.~Jerrold Franklin for correcting the notation in formula (\ref{eq:1}).


\end{document}